\documentclass{article}

\usepackage[utf8]{inputenc} 
\usepackage[T1]{fontenc}    
\usepackage{hyperref}       
\usepackage{url}            
\usepackage{booktabs}       
\usepackage{amsfonts}       
\usepackage{nicefrac}       
\usepackage{microtype}      
\usepackage{lipsum}
\usepackage{verbatim}
\usepackage{graphicx}
\usepackage{amsmath}
\usepackage{amssymb}
\usepackage{lineno}
\usepackage{authblk}
\usepackage{geometry}
\usepackage{graphicx}
\usepackage{caption}
\usepackage{ulem}
\usepackage{color}
	 \definecolor{darkred}{rgb}{1,0,0}
	 \definecolor{darkgreen}{rgb}{0,0.5,0}
	 \definecolor{darkblue}{rgb}{0,0,1}
  	 \definecolor{darkorange}{rgb}{1,0.7,0.2}
	 \definecolor{dark}{rgb}{0,0,0}

\title{Evolutionary dynamics of zero-determinant strategies in repeated multiplayer games}

\author[1]{Fang Chen}
\author[2]{Te Wu}
\author[1,3]{Long Wang \thanks{Corresponding author: Long Wang(longwang@pku.edu.cn)}}
\affil[1]{\footnotesize Center for Systems and Control, College of Engineering, Peking University, Beijing, China}
\affil[2]{Center for Complex Systems, Xidian University, Xi'an, China}
\affil[3]{Center for Multi-Agent Research, Institute for Artificial Intelligence, Peking University, Beijing, China}

\begin{document}
\maketitle

\begin{abstract}
Since Press and Dyson's ingenious discovery of ZD (zero-determinant) strategy in the repeated Prisoner's Dilemma game, several studies have confirmed the existence of ZD strategy in repeated multiplayer social dilemmas. However, few researches study the evolutionary performance of multiplayer ZD strategies, especially from a theoretical perspective. Here, we use a newly proposed state-clustering method to theoretically analyze the evolutionary dynamics of two representative ZD strategies: generous ZD strategies and extortionate ZD strategies. Apart from the competitions between the two strategies and some classical strategies, we consider two new settings for multiplayer ZD strategies: competitions in the whole ZD strategy space and competitions in the space of all memory-1 strategies. Besides, we investigate the influence of level of generosity and extortion on the evolutionary dynamics of generous and extortionate ZD, which was commonly ignored in previous studies. Theoretical results show players with limited generosity are at an advantageous place and extortioners extorting more severely hold their ground more readily. Our results may provide new insights into better understanding the evolutionary dynamics of ZD strategies in repeated multiplayer games.

\end{abstract}

\textbf{Keywords:} Evolutionary game, Direct reciprocity, Zero-determinant strategies, Repeated multiplayer games

\section{Introduction}
Direct reciprocity is a fundamental mechanism to promote the evolution of cooperation \cite{nowak2006five,barfuss2020caring,hilbe2017memory,su2019evolutionary,li2020evolution,nowak2004emergence,wu2017coevolutionary,wu2018coevolutionary,van2012emergence,akccay2018collapse,pinheiro2014evolution}. Recently, a class of strategies named zero-determinant (ZD) strategies in direct reciprocity \cite{press2012iterated} is of particular interest \cite{hilbe2015evolutionary,hilbe2013adaptive,pan2015zero,mamiya2020zero,adami2013evolutionary}. Such strategies, proposed by Press and Dyson, allow a player to enforce a linear relationship between his own payoff and his co-player's payoff, regardless of his co-player's strategy. A subset of ZD strategies consists of extortionate strategies, with which a player always gets more than his co-player does. It seems that an extortioner is always the winner, but it was proved that extortionate strategies are not evolutionarily stable unless for very small populations \cite{hilbe2013evolution}. Evolution leads to another subset of ZD strategies, so-called generous ZD \cite{stewart2013extortion}. Generous ZD strategies reward cooperation but punish defection mildly, leading to a cooperative state.

ZD strategies are not confined to repeated two-player games. Recently published studies have shown that ZD strategies also exist in repeated multiplayer games both with and without discounted payoffs \cite{hilbe2014cooperation,govaert2020zero}. Following the discoveries of ZD strategies in repeated multiplayer games, little is known about their evolutionary performance, especially theoretically. Questions yet to be answered include: which role do multiplayer ZD strategies play in the evolution of cooperation  \cite{boyd1988evolution,stewart2012extortion}? How does group size influence the evolutionary performance of multiplayer ZD strategies? The difficulty of dealing with these problems lies in the high complexity of calculating long-term payoffs \cite{hilbe2014cooperation}. To compute the long-term payoffs of an $n$-player repeated game, we need to calculate the stationary distribution of a Markov chain of dimension $2^n$. The dimension of the Markov chain increases exponentially with the number of players, making it impossible to compute long-term payoffs of large-scale repeated games, such as a repeated ten-player game. Recently, a state-clustering method \cite{chen2021stateclustering} significantly reduces the dimension of the Markov chain. Using this method, the computing complexity is reduced from $O(2^n)$ to $O(n^2)$, thus we can study the evolutionary dynamics in repeated multiplayer games.    

In this paper, we explore the evolutionary performance of ZD strategies using the state-clustering method. We begin our analysis by restricting the evolutionary competitions to the space of ZD strategies. We find the evolution eventually stabilizes at either generous strategies or extortionate strategies, depending on two key parameters capturing the essence of ZD strategies. Next, we explore the evolutionary dynamics between ZD strategies and some of the most important memory-1 strategies. Finally, we investigate the evolutionary performance of ZD strategies by measuring their abilities to resist invasion attempts and their abilities to invade other strategies. Analysis in ZD strategy space and memory-1 strategy space induces a consistent conclusion: less participants in a game and larger populations are conducive to the evolution of generous players but are detrimental to the prevalence of extortioners. The level of generosity and extortion of a ZD strategy also plays a vital role: a generous strategy with limited generosity dominates the population more easily while a extortioner extorting more severely prevails more readily. Our work provides an evolutionary perspective for multiplayer ZD strategies.

\section{Model}
\paragraph{Repeated multiplayer social dilemmas.}
Consider a repeated multiplayer game with $n$ players. In each round, players have two choices: cooperating (C) and defecting (D). The payoff of a player depends on his action and the number of cooperators among his co-players as Table \ref{table1} shows. If $k$ of $n-1$ co-players cooperate, a cooperator gets $a_k$ and a defector obtains $b_k$. We are mostly concerned with the social dilemma scenario which necessitates three characteristic sets of inequalities: (i) every player prefers his co-players to cooperate ($a_{k+1} \geq a_k$ and $b_{k+1} \geq b_{k}$); (ii) in a mixed group, a cooperator's payoff is always lower than the payoff of a defector ($b_{k+1}>a_k$); and (iii) mutual cooperation generates higher group benefit than mutual defection ($a_{n-1}>b_0$). It can be readily deduced that individuals prefer defection over cooperation while group payoff is maximized for all group members cooperating.

The public goods game is a typical social dilemma involving the interaction of multiplayer. In this game, each cooperator contributes a constant endowment $c>0$ to the common pool. Defectors contribute nothing. The total contribution is multiplied by a synergy factor $r$ $(1<r<n)$, and then evenly distributed to every player, regardless of his action. Thus, the payoff can be written as $a_k=r c(k+1)/n-c$ for a cooperator and $b_k=r c k/n$ for a defector.

\begin{table}[ht]
    \centering
    \begin{tabular}{c|cccccc}
    \hline
         \textbf{number of cooperators among co-players} & 0 & 1 & $\cdots$ & $k$ & $\cdots$ & $n-1$  \\ \hline
         cooperator's payoff & $a_0$ & $a_1$ & $\cdots$ & $a_k$ & $\cdots$ & $a_{n-1}$  \\ \hline
         defector's payoff & $b_0$ & $b_1$ & $\cdots$ & $b_k$ & $\cdots$ & $b_{n-1}$ \\
         \hline
    \end{tabular}
    \caption{\textbf{Payoffs of multiplayer games.} In a group with $n$ players, each player either cooperates or defects. The payoff of a player depends on his action and the number of cooperators among his co-players. If $k$ of $n-1$ co-players cooperate, a cooperator will get $a_k$ and a defector will obtain $b_k$.}
    \label{table1}
\end{table}

\paragraph{Zero-determinant strategies.} In repeated multiplayer social dilemmas, players can condition their actions on outcomes in previous rounds. In such settings, a strategy prescribes how to react given one of all possible histories so far. Generally speaking, with the folding of interactions, the set of feasible strategies expands exponentially. Here we limit our focus on the complete set of memory-1 strategies. Players using memory-1 strategies base their decisions just on the previous round. A memory-1 strategy can be written as a vector $\mathbf{p}=(p_{C0},\cdots,p_{Cn-1},p_{D0},\cdots,p_{Dn-1})$, where $p_{Ak}$ represents the probability a player cooperates after the round in which he has taken action $A$ ($A \in \{C, D\}$) and $k$ of $n-1$ co-players have cooperated. Repeat $\mathbf{p}^{Rep}$ is a typical memory-1 strategy. It simply repeats one's own move of the previous round and thus $p_{Ck}=1$ and $p_{Dk}=0$. 

We would like to analytically address the evolutionary dynamics using the state-clustering method of payoff computation (details in Appendix B). This newly discovered method can significantly reduce the dimension of the corresponding Markov chain used to derive payoffs in the limit of small mutation rates. Take $n=10$ as an example. For a small mutation, these 10 players are of at most two different strategies \cite{fudenberg2006imitation}. If we list all possible states consisting of each player's action, the Markov chain is of dimension $2^{10}$. However, using the state-clustering method, the dimension can be reduced from $2^{10}$ to $2(k+1)(n-k)$ with $k$ being the number of players of one strategy. This significant reduction shortens the time consumed on computing payoffs, especially when the procedures repeat. More importantly, it can rigorously derive payoffs for large groups which otherwise is impossible.

\begin{figure}[ht]
  \centering
  \includegraphics[width=2.5in]{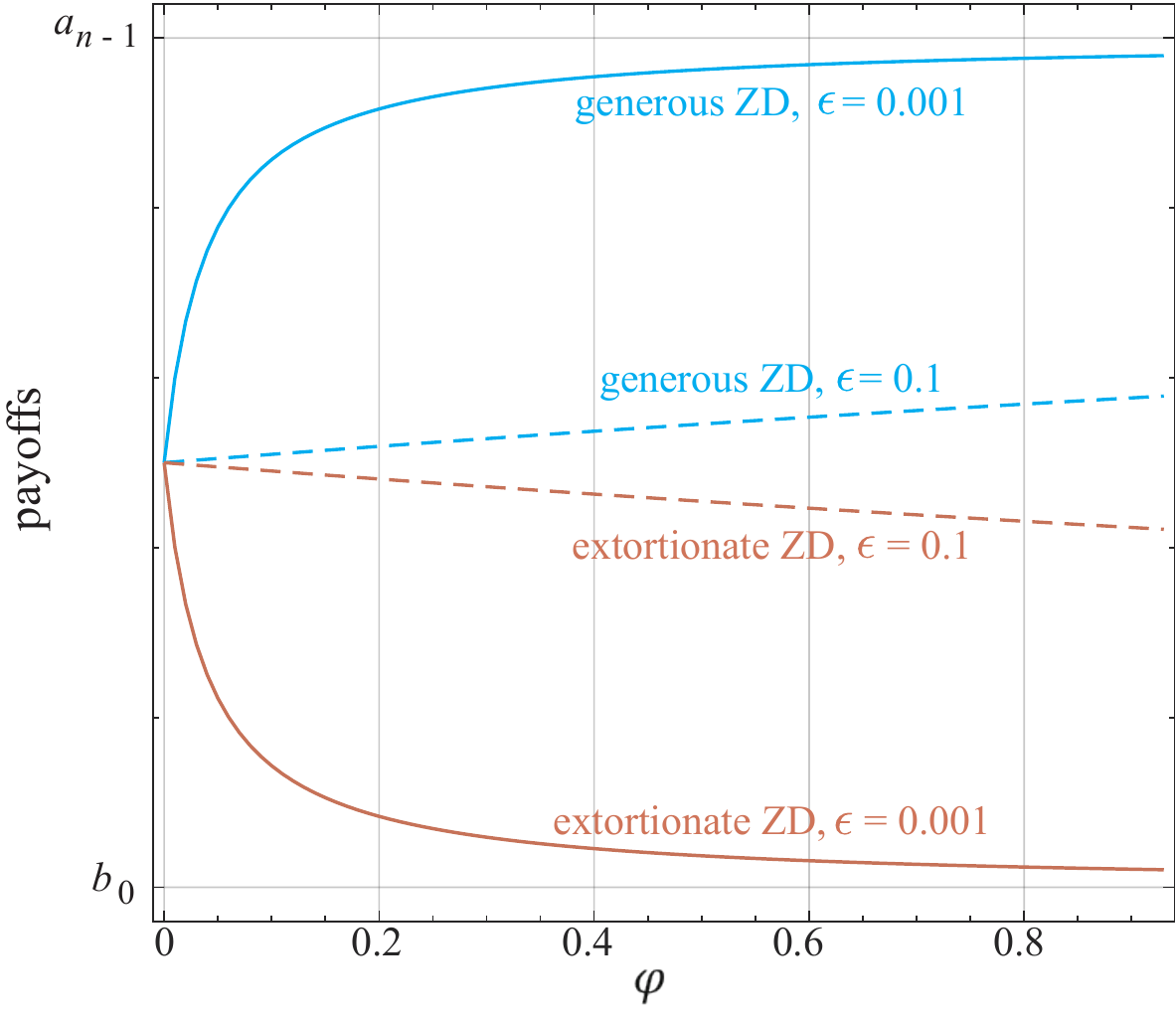}
  \caption{\textbf{The influence of $\varphi$.} Consider a group where all players apply the same generous (extortionate) ZD. If all players execute their actions exactly, each player obtains $a_{n-1}$ ($b_0$). If players' actions are subject to implementation errors, the effective strategy of a player with $\mathbf{p}$ becomes $(1-\epsilon)\mathbf{p}+\epsilon(1-\mathbf{p})$. $\varphi$ portrays the influence of implementation errors on the long-term payoff that a player gets in a group with all players applying the same ZD strategy. Great $\varphi$ keeps the long-term payoff near original value.}
  \label{fig:error}
\end{figure}

Recent discovery of a new class of memory-1 strategy, zero-determinant (ZD) strategies, has attracted great interest in the evolutionary dynamics of repeated games. A player using such ZD strategy can enforce a linear relationship between his payoff and his co-players' average payoff. Let $\pi_j$ denote the payoff of player $j$. Let
\begin{equation}
    \pi_{i}-l=w \sum_{j \neq i}\left(\pi_{j}-l\right)
    \label{equ:ZDrelationship}
\end{equation}
be the relationship player $i$ wants to enforce. 
According to Akin's lemma \cite{akin2016iterated}, player $i$ can enforce such relationship using a ZD strategy in the form of
\begin{equation}
    \mathbf{p} = \mathbf{p}^{Rep} + \varphi[\mathbf{S}^i - w\sum_{j \neq i} \mathbf{S}^j - \left(1-(n-1)w\right)l\mathbf{1}],
    \label{equ:ZDstrategy}
\end{equation}
where $\mathbf{S}^i=(S^i_{A,j})$, with $S^i_{C,j}=a_j$ and $S^i_{D,j}=b_j$, is the payoff vector of player $i$. $\mathbf{1}$ is a vector with all elements being 1. $l$ represents the baseline payoff of strategy \eqref{equ:ZDstrategy} since every player gets $l$ if all players apply the same ZD strategy. The parameter $w$ weights the degree to which player $i$ intents to extort his co-players'. Small $w$ means player $i$ almost obtains the same payoff as the average payoff of the group. $\varphi$ delimits the parameter range over which the ZD strategy is feasible. Assume players do not execute their actions exactly, but their actions are subject to implementation errors: when a player decides to cooperate (defect), he defects (cooperates) with probability $\epsilon$. Thus, the effective strategy of a player with $\mathbf{p}$ becomes $(1-\epsilon)\mathbf{p}+\epsilon(1-\mathbf{p})$. $\varphi$ also measures the degree of influence by implementation errors on the payoff that a player gets when every player in the group takes the same ZD strategy (Fig. \ref{fig:error}). Small $\varphi$ causes greater influence of implementation errors. 

Two subsets of ZD strategies receive considerable attention. One is extortionate ZD strategies, which set the baseline at $b_0$. The other subset is generous ZD strategies, setting the baseline payoff at $a_{n-1}$. Extortioners always get over the average payoff of a group, but yield the minimum payoff $b_0$ if all players take extortionate ZD. Generous players always get below the average, but lead to a cooperative state. 

\section{Results}
\paragraph{Evolutionary performance in the space of ZD strategies.}
We use adaptive dynamics \cite{zhang2013tale,ginsberg2019evolution,dieckmann1996dynamical,hilbe2013adaptive,cressman2012cooperation} to study the evolutionary performance of ZD strategies in the case where all players adopt ZD strategies. For rare mutations, the population is most of time monomorphic. Adaptive dynamics suggest the monomorphic population evolves in the direction of nearby population that leads to higher payoffs than the resident strategy. The canonical equations \cite{dieckmann1996dynamical} of adaptive dynamics are given by
\begin{equation*}
    \left\{
    \begin{aligned}
      \frac{d w}{d t}&=\left.\frac{\partial \pi\left(l^{\prime},w^{\prime}; l,w\right)}{\partial w^{\prime}}\right|_{w^{\prime}=w, l^{\prime}=l}\\
      \frac{d l}{d t}&=\left.\frac{\partial \pi\left(l^{\prime},w^{\prime}; l,w\right)}{\partial l^{\prime}}\right|_{w^{\prime}=w, l^{\prime}=l}
    \end{aligned}
    \right.,
\end{equation*}
where $\pi\left(l',w'; l,w\right)$ is the payoff of the player who deviates from $(l,w,\varphi)$ to $(l',w',\varphi')$ in a group with all players applying $(l,w,\varphi)$. We can get $\pi\left(l',w'; l,w\right)$ as (details in Appendix C)
\begin{equation*}
	\pi\left(l^{\prime}, w^{\prime};l, w\right)=\frac{\left[1-(n-1) w^{\prime}\right][1-(n-2) w] l^{\prime}+(n-1) w^{\prime}[1-(n-1) w] l}{1-(n-2) w-(n-1) w w^{\prime}}.
\end{equation*}
Thus, the canonical equations can be written as
\begin{align}
    \frac{d w}{d t}&=\left.\frac{\partial \pi\left(l^{\prime},w^{\prime}; l,w\right)}{\partial w^{\prime}}\right|_{w^{\prime}=w, l^{\prime}=l}=0 \label{equ:adaptivedynamics1} \\
    \frac{d l}{d t}&=\left.\frac{\partial \pi\left(l^{\prime},w^{\prime}; l,w\right)}{\partial l^{\prime}}\right|_{w^{\prime}=w, l^{\prime}=l}=\frac{1-(n-2) w}{1+w}.\label{equ:adaptivedynamics2}
\end{align} 
Eq. \eqref{equ:adaptivedynamics1} implies the weight $w$ stays constant under adaptive dynamics. It should be noted that $w$ determines the evolutionary result of baseline payoff $l$. For $w>1/(n-2)$, the baseline payoff $l$ decreases over time. Eventually, the population yields the minimum payoff $b_0$ and extortioners pervade the whole population. On the contrary, for $w<1/(n-2)$, the baseline payoff $l$ increases over time and eventually stabilizes at $a_{n-1}$, the possibly maximal payoff. This means generous players take over the whole population. We can thus claim that small weight favors the evolution of generous ZD strategies. This is intuitively understandable. A generous player with a small $w$ gets payoff below the average payoff of all other group members, but his strategy setting just allows a limited payoff difference. On the other hand, generous players reciprocate each other, totally overwhelming their loss of being exploited by highly defective players. As a result, players will stick to generous strategies and not deviate.

\begin{figure}
  \centering
  \includegraphics[scale=0.55]{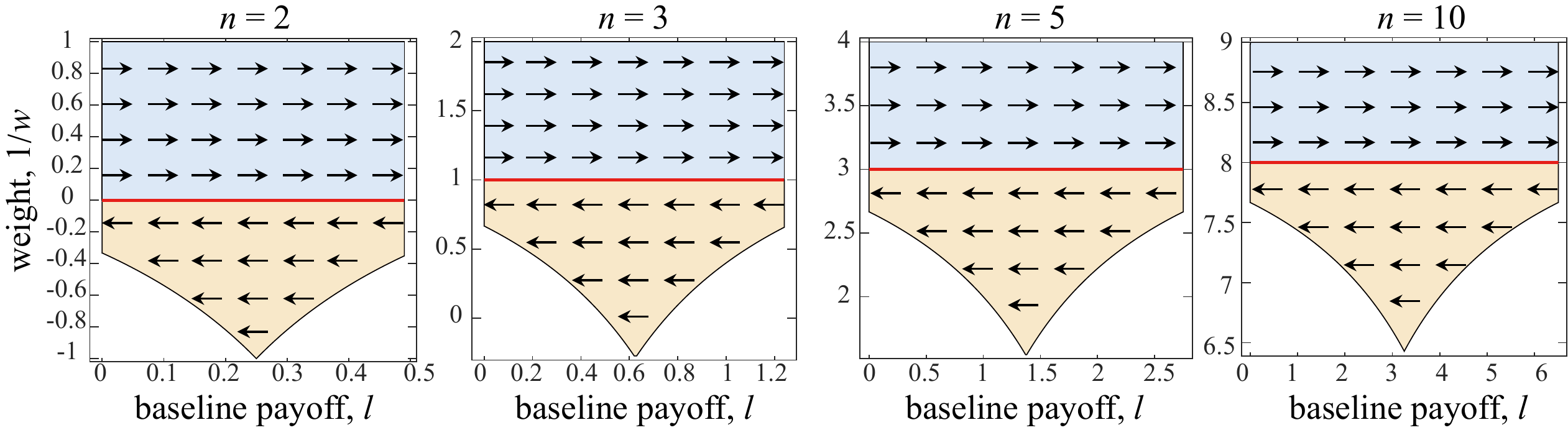}
  \caption{\textbf{Evolutionary performance in the space of ZD strategies.} We use adaptive dynamics to explore the evolutionary performance of ZD strategies in the whole ZD strategy space. The red line represents $w=1/(n-2)$. For $w<1/(n-2)$, the population is stabilized at the state with all players being generous. For $w>1/(n-2)$, extortionate strategies are the evolutionary winner. Therefore, a small weight is conducive for the evolution of generous players and a large weight contributes to the prevalence of extortioners. As the group size increases, the attraction domain of generous ZD strategies (blue range) decreases, indicating large group size hinders the evolution of generosity. Parameters: $\delta=r/n=0.75$ and $c=1$.}
  \label{fig:adaptive}
\end{figure}

ZD strategies with intermediate baseline payoff, $l \in (b_0,a_{n-1})$, are not evolutionary stable strategies under adaptive dynamics \cite{foster1990stochastic,nowak2006evolutionary}. Thus, the population is composed of either generous ZD strategies or extortionate ZD strategies. As the group size increases, the attraction domain of generous ZD strategies decreases while that of extortionate strategies increases (Fig. \ref{fig:adaptive}), indicating that large group size hinders the evolution of generosity.

\begin{figure}[h!]
  \centering
  \includegraphics[width=5.7in]{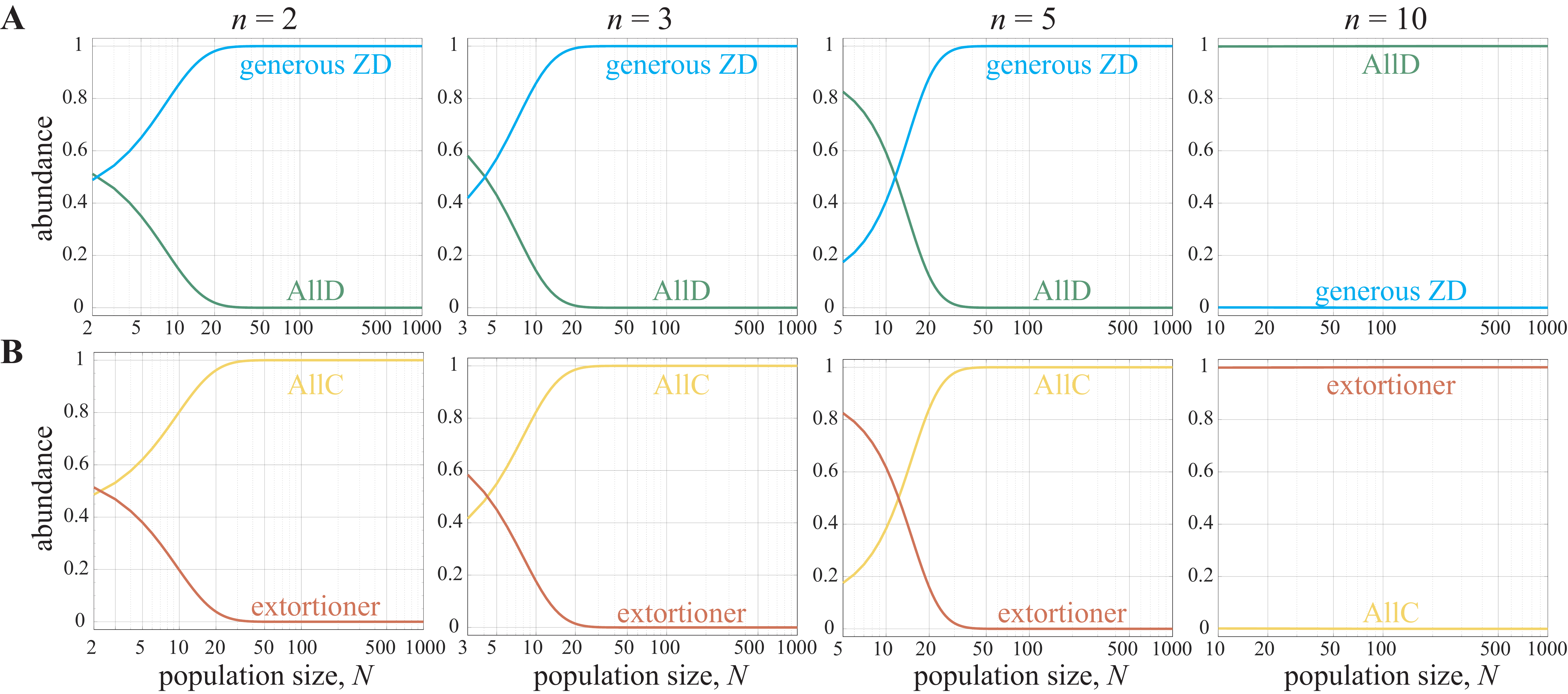}
  \caption{\textbf{Competitions between ZD strategies and classical memory-1 strategies.} (A) Competitions between generous ZD and AllD. In small groups, generous strategies are the evolutionary outcome in sufficiently large populations. In large groups, generous strategies no longer dominate in any population, implying large groups hinder the evolution of generosity. (B) Competitions between extortioners and AllC players. Extortioners can only prevail in very small populations if the group size is small. Yet, extortioners occupy the population with arbitrary size if groups are large. The conclusion that large groups are conducive to the evolution of extortioners can be obtained from this observation. Parameters: $\delta=r/n=0.75$, $c=1$, $\epsilon=0.001$ and $w=1.1/(n-1)$.  }
  \label{fig:competitions}
\end{figure}

\paragraph{Competitions between ZD strategies and classical memory-1 strategies.} Consider a finite population with $N$ players and evolving following a mutation-selection process. In each time step, a player $X$ is randomly chosen to update the strategy. With probability $\mu>0$, mutations happen. Player $X$ would explore a strategy drawn from a uniform distribution over the space of memory-1 strategies. With probability $1-\mu$, imitation happens. Player $X$ imitates the strategy of another randomly chosen player $Y$. Denote by $P_X$ the average payoff of $X$ and by $P_Y$ the average payoff of $Y$. $X$ switches to $Y$'s strategy with probability $1/[1+e^{s(P_X-P_Y)}]$, where $s$ is the selection intensity. Great $s$ indicates payoffs contribute more to fitness. Assume the mutation rate $\mu$ is low. Thus, the population is almost monomorphic all the time. Such stochastic process yields a steady-state distribution of strategies.

Given that ZD strategies with intermediate baseline payoff, $l\in(b_0,a_{n-1})$, are not evolutionarily stable, we mainly focus on generous and extortionate ZD strategies. Consider two representative pairwise competition dynamics, generous players competing with always defection (AllD) and extortioners competing with always cooperation (AllC). When generous players compete with AllD players, Fig. \ref{fig:competitions}A shows that the threshold population size required for generous strategies to be dominant increases as the group size rises from 2 to 3 to 5. Before the threshold, the abundance of generous strategy also declines for the same population size as the group size increases. For larger group of size 10, generous strategy is always eliminated as the population size varies from 10 to 1000. In fact, increasing group size does reinforce the mutual breed of generous ZD strategies. It is implied in Fig. \ref{fig:competitions}A that the force of exploitation outperforms the force of the mutual breed, making it more difficult for generous ZD strategies to be selected. This implication is further confirmed in another competition settings: AllC competing with extortioners. AllC players are highly cooperative. Extortioners are highly defective. AllC players are more likely to reciprocate mutually with the expansion of group size, which can also lead to severe exploitation of extortioners over AllC players. Similar evolutionary outcomes are observed as generous players competing with AllD players, once again corroborating that the expansion of group size benefits defective strategies more than cooperative strategies.

\begin{figure}[h!]
  \centering
  \includegraphics[width=5.7in]{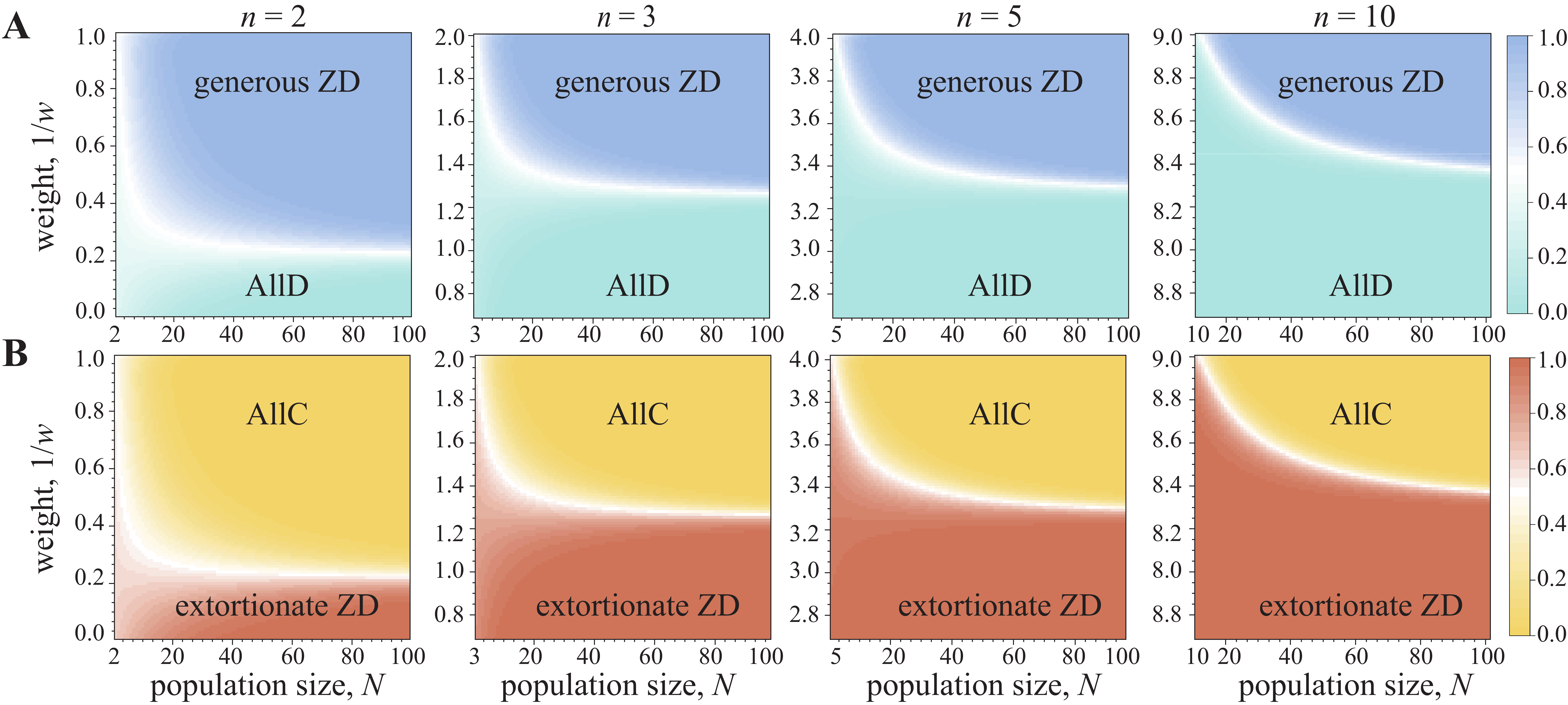}
  \caption{\textbf{The influence of weights on the evolution of ZD strategies.} (A) The influence of weights on the fraction of generous ZD. (B) The influence of weights on the fraction of extortioners. As the weight increases, defective strategies (AllD or extortionate ZD) take over the population most of the time. The range of weight favoring the evolution of generous ZD and AllC narrows as the group size increases. Generous ZD strategies and AllC lead to cooperative states. Large groups hinder the evolution of cooperative strategies. Parameters are $\delta=r/n=0.75$, $c=1$, $\epsilon=0.001$ and $N=100$.}
  \label{fig:parameter}
\end{figure}

We further explore the influence of the weight $w$ on the evolution of ZD strategies (Fig. \ref{fig:parameter}). We get qualitatively similar results as by adaptive dynamics. Small weights are conducive to the evolution of generous ZD strategies. As the group size increases, the parameter range of $w$ favoring generous ZD strategies contracts, indicating that large groups impede the evolution of generosity. Differently, larger group size favors the evolution of extortionate ZD strategies. For small group size, extortioners who extort very severely can also be selected and hold their ground readily. As group size increases, to survive the evolutionary competition, extortioners can extort highly cooperative opponents less severely, illustrated by the compressed area of $w$ favoring extortion as the group size increases.

\paragraph{Evolutionary performance of ZD strategies in the space of memory-1 strategies.} We study the evolutionary performance of ZD strategies in the memory-1 strategy space using the evolutionary stochastic dynamics of Imhof and Nowak \cite{imhof2010stochastic}. According to Imhof and Nowak, a successful strategy is a combination of two factors: (i) it depends on the ability of the strategy to resist invasion attempts, and (ii) it depends on the ability of the strategy to invade other (successful) strategies. To investigate the evolutionary performance of generous and extortionate strategies, we respectively test their these two abilities. To explore the ability of a strategy to resist invasion attempts, we initialize the population at the strategy. Then, we introduce a randomly drawn memory-1 strategy to the population. The mutant either reaches fixation or is wiped out. If the mutant is wiped out, another randomly drawn memory-1 strategy is introduced to the population as a new mutant. The algorithm enters the next cycle. The algorithm continues until the first mutant reaches fixation. We record the number of mutants it takes for fixation to occur. If payoffs contribute nothing to fitness, the well-known neutral drift \cite{nowak2005evolution}, it takes $N$ mutants on average to reach fixation. A strategy is favored by selection if it resists more than $N$ mutants. 

\begin{figure}
  \centering
  \includegraphics[width=5.7in]{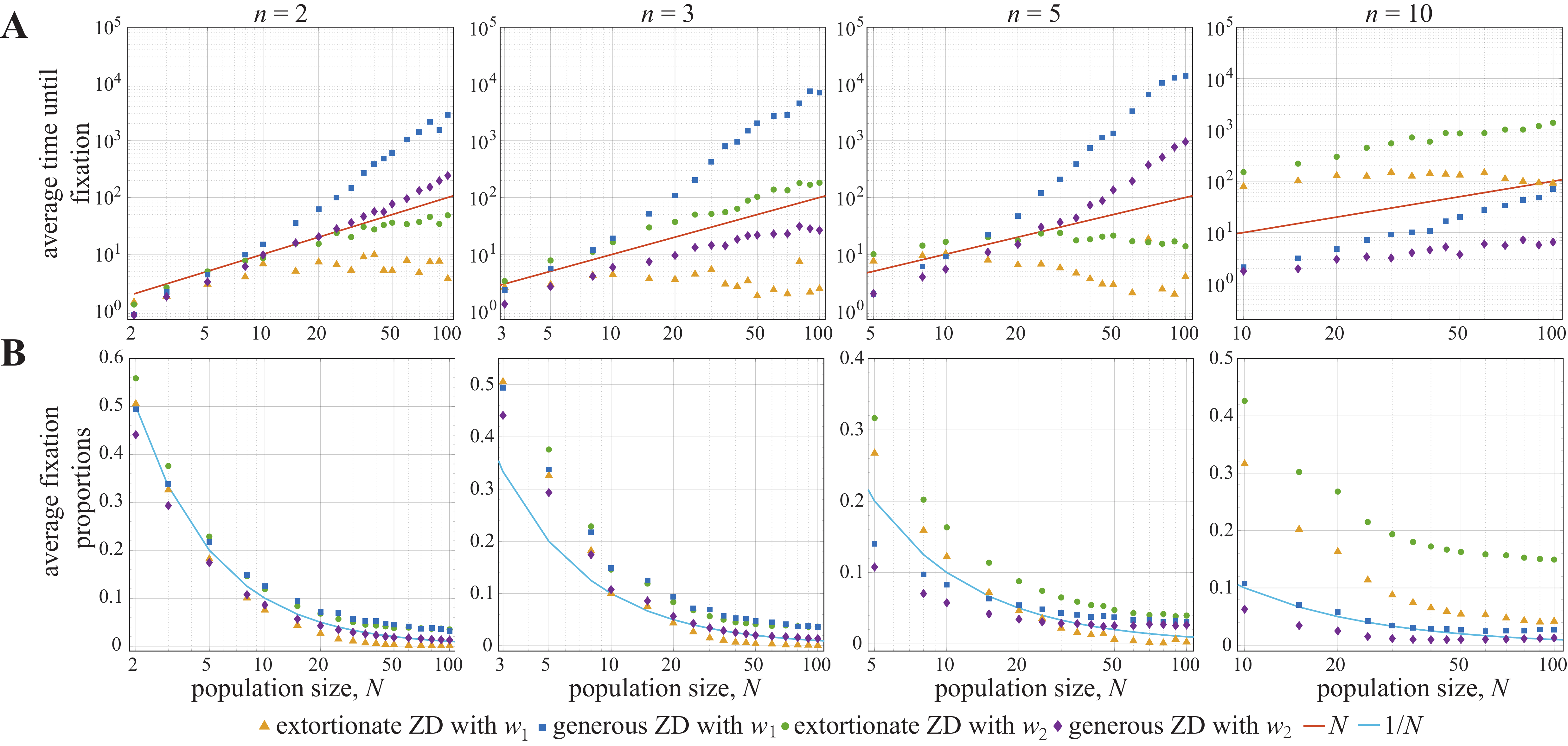}
  \caption{\textbf{The evolutionary performance of ZD strategies in memory-1 strategy space.} We use stochastic evolutionary dynamics to explore the evolutionary performance of ZD strategies from two sides: the ability of ZD strategies resisting invasion attempts (A) and the ability to invade other strategies (B). We find that extortioners are supported by selection to resist invasion attempts and to invade other strategies in large groups. On the contrary, generous players can resist more invasion attempts and invade more populations for small groups. We consider two different weights, $w_1$ and $w_2$ ($w_1<w_2$). In any population, generous strategies with small weights resist more invasion attempts and invade more strategies while extortioners with large weights have stronger ability to resist mutants and to invade other strategies. Parameters: $\delta=r/n=0.75$, $c=1$ and $\epsilon=0.001$. $w_1=1.1$ and $w_2=5$ if $n=2$. $w_1=0.55$ and $w_2=1$ if $n=3$. $w_1=0.275$ and $w_2=0.3$ if $n=5$. $w_1=0.122$ and $w_2=0.13$ if $n=10$. }
  \label{fig:SED}
\end{figure}

In repeated multiplayer games with 2, 3 and 5 players, generous players are favored by natural selection as always more than $N$ (=the population size) invasion attempts are needed to wipe out generous players for medium or large populations (Fig. \ref{fig:SED}A). However, in repeated games with 10 players, it needs less than $N$ invasion attempts to fixate in populations of generous players. Compare to neutral drift, extortioners are more easily to be invaded as the population size increases, indicating large groups are not beneficial for extortioners to resist invasion attempts. Even though, when group size reaches 10, extortioners are favored by selection in any population. We investigate generous ZD and extortionate ZD each with two different weights, $w_1$ and $w_2$ ($w_1<w_2$). Generous players with small weights can resist more mutations. Extortioners resist more mutants when they extort more severely. 

To investigate the ability of a strategy $\mathbf{p}$ to invade other strategies, we randomly draw $10^5$ strategies from the space of memory-1 strategies. Then, let $\mathbf{p}$ as mutant invade these strategies. The mutant takes over the population with probability $\left(1+\sum_{i=1}^N\prod_{k=0}^i \exp[s(P_R(k)-P_M(k))]\right)^{-1}$, where $P_M(k)$ and $P_R(k)$ respectively denote payoffs of mutants and residents if there are $k$ mutants in the population. We count the average proportion of these $10^5$ strategies the mutant can take over. If such average fixation proportion of a strategy is over $1/N$, we say the strategy is supported by selection to invade other strategies.

Results show that the population size and the group size have qualitatively similar effects on the ability to invade other strategies as on the ability to resist invasion attempts (Fig. \ref{fig:SED}B). Weights play a vital role in determining the evolutionary fates of generous ZD strategies and extortionate strategies. Generous strategies with small weights are more favored by selection, especially in large populations. Extortioners extorting to a large degree always invade more than one out of $N$ randomly assigned strategies, which is more successful than under neutral drift.

\section{Discussion}
Zero-determinant (ZD) strategies are of particular interest due to their unusual degree of control over opponents' long-term payoffs \cite{taha2020zero,ichinose2018zero,hilbe2015partners}. Following the original work on the ZD strategy in repeated prisoner's dilemma, several studies have independently shown that ZD strategies also exist in repeated multiplayer games. However, in repeated multiplayer games, little is known, especially on the theoretical front, about the evolutionary performance of multiplayer ZD strategies. This is mainly due to the high complexity of computing payoffs as pointed out in \cite{hilbe2014cooperation}: ``the mathematics of repeated $n$-player dilemmas seems to be more intricate, and numerical investigations are impeded because the time to compute payoffs increases exponentially in the number of players.'' We have recently developed a state-clustering method which could reduce the computing complexity from $O(2^n)$ to $O(n^2)$ \cite{chen2021stateclustering}. Using this efficient method, we have explored the evolutionary performance of ZD strategies in three representative competition settings: competitions between all ZD strategies, ZD strategies competing with classical memory-1 strategies, and evolutionary race in the whole space of memory-1 strategies. Results consistently in three settings show that: large populations, small groups each favor the evolution of generous strategies while hinder the prevalence of extortionate strategies. Moreover, ZD strategies that are generous but not too generous support their prevalence. On the contrary, ZD strategies that are extortionate but to a limited degree put themselves in a disadvantageous place in the evolutionary race. 

For iterated prisoner's dilemma, Hilbe et al. has shown that extortioners are eliminated as long as the population size reaches a threshold, even if the competition settings are as simple as just containing AllD, WSLS and extortionate ZD, with or without AllC \cite{hilbe2013evolution}. Instead of finding the evolutionary stable strategies, which is very demanding or even impossible, Steward et al. have proposed the concept of evolutionary robust strategy \cite{stewart2013extortion}. They have also given the condition, $1\le w \le (2N+1)/(N+1)$, for the subset of generous ZD strategies to be evolutionary robust. It follows that large populations allow a wider parameter range of admitting robust generous ZD strategies. Our results demonstrate that large populations still favor the robustness of generous ZD strategies in repeated multiplayer games. In comparison with neutral drift, generous players are hard to be invaded by other strategies and able to invade numerous homogeneous populations each consisting of one type of different strategies. In contrast, extortioners are more easily to be invaded and less able to invade other strategies.

In iterated prisoner's dilemma, extortioners are selected only in very small populations \cite{hilbe2013evolution}. However, in repeated multiplayer with large groups, extortionate ZD strategies are always selected for any population size. On the contrary, generous ZD strategies enjoy a great abundance in reasonably large populations in iterated prisoner's dilemma. As group size increases, the competence of generous ZD strategies is discounted. For group size increasing from 2 to 3 to 5, the population size required for generous players to be dominant increases from 3 to 5 to 12. For the group as large as 10, generous ZD strategies are nearly wiped out. Large groups allow extortioners to exploit more players at the same time. On the other hand, generous players are likely to be severely exploited by defective players in large groups. These two factors lead to selection favoring extortionate strategies over generous ZD strategies, an observation consistent with ones reported in \cite{chen2021stateclustering} and \cite{boyd1988evolution}. 

The payoff weight $w$ plays a crucial role in determining the evolutionary fate of ZD strategies. Small $w$ means the player using ZD strategy obtains payoff very close to the average payoff of all other group members. It was found that generous ZD in repeated prisoner's dilemma games is evolutionary robust if $1\le w \le (2N+1)/(N+1)$ \cite{stewart2013extortion}. We have obtained qualitatively similar results for repeated multiplayer games. Small $w$ is beneficial for the evolution of generous ZD strategies. Large $w$ disfavors the evolution of generous ZD but puts extortionate ZD strategies at an advantageous place. We hope that our study would shed light on better understanding evolutionary dynamics in repeated multiplayer games, especially theoretically.

\section*{Appendix A: ZD strategies}\label{sec:zd}
In this part, we prove player $i$ adopting a memory-1 strategy of the form
\begin{equation}
    \mathbf{p} = \mathbf{p}^{Rep} + \varphi[\mathbf{S}^i - w \sum_{j \neq i}\mathbf{S}^j - \left(1-(n-1)w\right)l\mathbf{1}].
    \label{Aequ:ZDstrategy}
\end{equation}
can enforce a linear relationship between his and his co-players' payoffs as
\begin{equation}
   \pi_{i}-l=w \sum_{j \neq i}\left(\pi_{j}-l\right).
    \label{Aequ:ZDrelationship} 
\end{equation}
These strategies of form \eqref{Aequ:ZDstrategy} are so-called zero-determinant (ZD) strategies.

Let $v_{Ak}(t)$ denote the probability that the focal player takes action $A$ ($A\in\{C,D\}$) and $k$ of $n-1$ co-players cooperate at time $t$. Collect these probabilities as a vector $\mathbf{v}(t)$. Suppose the repeated game is played for infinitely many rounds. The mean distribution of $\mathbf{v}(t)$ is given by $\mathbf{v}= \lim_{t \rightarrow \infty} [\mathbf{v}(1)+\cdots+\mathbf{v}(t)]/t$. The expected payoff of player $i$ can be written as $\pi_i=\mathbf{S}^i \cdot \mathbf{v}$.

Denote by $q_C(t)$ the probability the focal player cooperates at time $t$. Thus, $q_C(t)$ can be written as $q_C(t)=\mathbf{p}^{Rep}\cdot \mathbf{v}(t)$. Suppose the focal player applies $\mathbf{p}$. $q_C(t+1)$ can be written as $q_C(t+1)=\mathbf{p} \cdot \mathbf{v}(t)$. It follows $u_C(t)=:q_C(t+1)-q_C(t)=(\mathbf{p}-\mathbf{p}^{Rep}) \cdot \mathbf{v}(t)$. Summing $u_C(t)$ from 1 to $t$, then dividing by $t$, we get $(\mathbf{p}-\mathbf{p}^{Rep}) \cdot [\mathbf{v}(1)+\cdots+\mathbf{v}(t)]/t = (q_C(t)-q_C(1))/t$, which has absolute value at most $1/t$. By taking the limit $t \to \infty$, it yields
\begin{equation}
    (\mathbf{p}-\mathbf{p}^{Rep})\cdot\mathbf{v}=0.
    \label{Aequ:akin}
\end{equation}
Substituting \eqref{Aequ:ZDstrategy} to \eqref{Aequ:akin}, we obtain \eqref{Aequ:ZDrelationship}.

\section*{Appendix B: Payoffs in memory-1 strategy space} \label{sec:payoffs_m1} 
In this section, we provide a detailed deduction of calculating payoffs using the state-clustering method. Assume mutations are rare such that there are at most two different strategies in a population. Let $\mathbf{p}$ and $\mathbf{q}$ denote the effective strategies when players execute these two strategies. Here, $\mathbf{p}=(p_{C0},...,p_{Cn-1},p_{D0},...,p_{Dn-1})$ and $\mathbf{q}=(q_{C0},...,q_{Cn-1},q_{D0},...,q_{Dn-1})$. Suppose $k$ co-players of the focal player apply $\mathbf{p}$ and $n-k-1$ co-players adopt $\mathbf{q}$. 

The repeated multiplayer game can be modelled by a Markov chain. Let $C_x^{\mathbf{p}}$ denote the state where $x$ $\mathbf{p}$ players cooperate and let $C_y^{\mathbf{q}}$ denote the state where $y$ $\mathbf{q}$ players cooperate. The state space of the Markov chain can be written as $\mathcal{S}=\{A C_x^{\mathbf{p}} C_y^{\mathbf{q}}|A\in\{C,D\}, x\in\{0,...,k\}, y\in\{0,...,n-k-1\}\}$. In the following, we calculate the transition probability from $A C_x^{\mathbf{p}} C_y^{\mathbf{q}}$ to $A' C_{x'}^{\mathbf{p}} C_{y'}^{\mathbf{q}}$. For convenience, we first introduce a function $f$ with $f(C)=1$ and $f(D)=0$. Assume the focal player adopts $\mathbf{p}$. At the state $A C_x^{\mathbf{p}} C_y^{\mathbf{q}}$, the focal player would cooperate in the next round with probability $p_{A(x+y)}$. Thus, the focal player switches his action from $A$ to $A'$ with probability 
\begin{equation}
    |1-f(A')-p_{A(x+y)}|.
    \label{equ:tranprofocal}
\end{equation} 
The $\mathbf{p}$ ($\mathbf{q}$) players who cooperate in the previous round would cooperate with probability $p_{C(x+y-1+f(A))}$ ($q_{C(x+y-1+f(A))}$). The $\mathbf{p}$ ($\mathbf{q}$) players who defect in the previous round would cooperate with probability $p_{D(x+y+f(A))}$ ($q_{D(x+y+f(A))}$). Therefore, the probability that the number of cooperators among $\mathbf{p}$ players changes from $x$ to $x'$ is
\begin{equation}
    \begin{aligned}
    \sum_{j=0}^{x^{\prime}} &\tbinom{x}{j}p_{C(x+y-1+f(A))}^{j}\left(1-p_{C(x+y-1+f(A))}\right)^{x-j} \\
    & \tbinom{k-x}{x'-j} p_{D(x+y+f(A))}^{x^{\prime}-j}\left(1-p_{D(x+y+f(A))}\right)^{(k-x)-\left(x^{\prime}-j\right)}.\label{equ:tranprop}
    \end{aligned}
\end{equation}
The probability that the number of cooperators among $\mathbf{q}$ players changes from $y$ to $y'$ is
\begin{equation}
    \begin{aligned}
    \sum_{j=0}^{y^{\prime}} &\tbinom{y}{j} q_{C(x+y-1+f(A))}^{j}\left(1-q_{C(x+y-1+f(A))}\right)^{y-j}\\
    &\tbinom{n-k-1-y}{y'- j} q_{D(x+y+f(A))}^{y^{\prime}-j}\left(1-q_{D(x+y+f(A))}\right)^{(n-1-k-y)-\left(y^{\prime}-j\right)}.
    \label{equ:tranproq}
    \end{aligned}
\end{equation}
The transition probability from $A C_x^{\mathbf{p}} C_y^{\mathbf{q}}$ to $A' C_{x'}^{\mathbf{p}} C_{y'}^{\mathbf{q}}$ is the product of Eq. \eqref{equ:tranprofocal}, Eq. \eqref{equ:tranprop} and Eq. \eqref{equ:tranproq}. 

Collect all transition probabilities between combinations of states in $\mathcal{S}$ as a matrix $\mathbf{M}$. Given that players can't execute their actions exactly, the transition matrix $\mathbf{M}$ is stochastic and primitive. Thus, $\mathbf{M}$ has a unique left eigenvector $\mathbf{v}$ corresponding to eigenvalue 1. Each element of $\mathbf{v}$, $v_{A C_x^{\mathbf{p}} C_y^{\mathbf{q}}}$, represents the probability that the focal player finds him at the state $A C_x^{\mathbf{p}} C_y^{\mathbf{q}}$. Thus, the payoff of the focal player is
\begin{equation}
    \pi = \sum_{x,y} v_{C C_{x}^{\mathbf{p}} C_y^{\mathbf{q}}} a_{x+y} + \sum_{x,y} v_{D C_{x}^{\mathbf{p}} C_y^{\mathbf{q}}} b_{x+y}.
  \label{equ:payoff}
\end{equation}

\section*{Appendix C: Payoffs under adaptive dynamics} \label{sec:payoffs}
Suppose in a group where all players apply ZD strategy $(l,w,\varphi)$, a player $X$ deviates to a nearby ZD strategy $(l',w',\varphi')$. Denote by $\pi(l,w;l',w')$ and $\pi(l',w';l,w)$ the payoffs of the resident and $X$, respectively. A resident can enforce a linear relationship as
\begin{equation}
    \pi(l,w;l',w')-l = w[(n-2)(\pi(l,w;l',w')-l)+(\pi(l',w';l,w)-l)],
    \label{Aequ:rrelation}
\end{equation}
The player $X$ can enforce a relationship as
\begin{equation}
    \pi(l',w';l,w)-l' = w'(n-1)(\pi(l,w;l',w')-l').
    \label{Aequ:mrelation}
\end{equation}
Solving simultaneous equations of \eqref{Aequ:rrelation} and \eqref{Aequ:mrelation}, we get
\begin{align}
    \pi(l,w;l',w')&=\frac{\left[1-(n-1) w^{\prime}\right] w l^{\prime}+[1-(n-1) w] l}{1-(n-2) w-(n-1) w w^{\prime}} \\
    \pi(l',w';l,w)&=\frac{\left[1-(n-1) w^{\prime}\right][1-(n-2) w] l^{\prime}+(n-1) w^{\prime}[1-(n-1) w] l}{1-(n-2) w-(n-1) w w^{\prime}}.
\end{align}

\paragraph{Acknowledgment}
We acknowledge support from the National Natural Science Foundation of China (NSFC 62036002) and PKU-Baidu Fund (2020BD017).

\normalem
\bibliographystyle{unsrt}

\end{document}